\documentclass[10pt,letterpaper]{article}
\usepackage[top=0.85in,left=2.75in,footskip=0.75in]{geometry}

\usepackage{changepage} 
\usepackage[utf8]{inputenc} 
\usepackage{textcomp,marvosym} 
\usepackage{fixltx2e} 
\usepackage{amsmath,amssymb} 
\usepackage{cite} 
\usepackage{nameref,hyperref} 
\usepackage{microtype} 

\DisableLigatures[f]{encoding = *, family = * }
\usepackage{rotating} 

\raggedright
\usepackage[aboveskip=1pt,labelfont=bf,labelsep=period,justification=raggedright,singlelinecheck=off]{caption}

\makeatletter
\renewcommand{\@biblabel}[1]{\quad#1.}
\makeatother

\date{}

\usepackage{lastpage,fancyhdr,graphicx}
\usepackage{epstopdf}
\fancyheadoffset[L]{2.25in}
\fancyfootoffset[L]{2.25in}
\begin{document}
\vspace*{0.35in}

\begin{flushleft}
{\Large
\textbf\newline{Spatio-temporal analysis of micro economic activities in Rome reveals patterns of mixed-use urban evolution}
}
\newline
\\
Alessandro Fiasconaro\textsuperscript{1,*},
Emanuele Strano\textsuperscript{2},
Vincenzo Nicosia\textsuperscript{1},
Sergio Porta\textsuperscript{3},
Vito Latora\textsuperscript{1,4}
\\
\bigskip
\bf{1} School of Mathematical Sciences, Queen Mary University
 of London, Mile End Road, E14NS London, UK
\\
\bf{2} Laboratory of Geographic Information Systems (LASIG),
  School of Architecture, Civil and Environmental Engineering (ENAC),
  Ecole Polytechnique F\'ed\'erale de Lausanne (EPFL), Switzerland
\\
\bf{3} Urban Design Studies Unit, University of Strathclyde,
  Glasgow, UK
\\
\bf{4} Dipartimento di Fisica ed Astronomia,
Universit\`a di Catania and INFN, I-95123 Catania, Italy
\\
\bigskip

%
%





* A.Fiasconaro@qmul.ac.uk

\end{flushleft}

\section*{Abstract}
Understanding urban growth is one with understanding how society
evolves to satisfy the needs of its individuals in sharing a
common space and adapting to the territory.  We propose here a
quantitative analysis of the historical development of a large
urban area by investigating the spatial distribution and the age
of commercial activities in the whole city of Rome. We find that
the age of activities of various categories presents a very
interesting double exponential trend, with a transition possibly
related to the long-term economical effects determined by the oil
crisis of the Seventies. The diversification of commercial
categories, studied through various measures of entropy, shows,
among other interesting features, a saturating behaviour with the
density of activities. Moreover, different couples of commercial
categories exhibit over the years a tendency to attract in space.
Our results demonstrate that the spatio-temporal distribution of
commercial activities can provide important insights on the
urbanisation processes at work, revealing specific and not trivial
socio-economical dynamics, as the presence of crisis periods and
expansion trends, and contributing to the characterisation of the
maturity of urban areas.


\section*{Introduction}
In present urbanism the idea that cities are mostly and
essentially condenser of social and economic activities is
popularly associated with Jane Jacobs' call for more compact urban
environments around socially inclusive open spaces
\cite{1961Jacobs}. It is today well known that cities emerge and
grow because density pays off, and does that at a pace that
largely off-sets that of its negatives
\cite{2013Batty,2009Batty,2007Glaeser,2007Bettencourt}. The first step in
understanding cities is therefore examining why activities are
concentrated in a few places. Certain activities in fact exhibit
an increasing return-to-scale, meaning that they profit
proportionally more, or cost proportionally less, than the spatial
growth of the city, and this is regarded as the driving force
behind the growth of cities \cite{2005OFlaherty}. However, what
makes cities great in expanding the benefit of concentration is
not just the economy of scale per se, but that whatever you do
occurs in a place where a number of other things also occur nearby
at the same time. Those nearby activities may belong to the same
industry ({\em
  localization economies}) or to a variety of different industries
({\em urbanization economies}). Understanding whether localization or
urbanization economies are the reason why cities grow, and under what
circumstances, is obviously very important to determine policies
towards increasing specialization or rather, on the other hand,
increasing heterogeneity.

Economic activities and the way they tend to locate in cities have
been investigated from the point of view of spatial economics,
looking at the equilibrium in space between factors like
customers, suppliers and transport costs that would determine the
optimal location of activities; this stream of studies, that draws
back to classics written as early as in the late XIXth-early XXth
century \cite{1933Christaller,1882Launhardt,1929Weber}, has paved
the way for an intense post-war period of progressive regional and
urban analysis \cite{2000Wilson}, which now struggles to make its
way into a globalized world dominated by uncertainty, expanding
urban poverty and the increasing fragmentation of decision making
systems, including urban planning systems.

A more recent area of scientific investigation of urban economics
has shifted the focus from business and employment only to retail,
acknowledging the fundamental role of the sector in processes of
urbanization and taking advantage of new potentials in Geographic
Information Systems (GIS) modeling \cite{2005Arentze}. This stream
of research refers directly to studies in marketing and business
organization, largely a response to industry's quest to develop
scientific models of geo-located advantage with regards to
potential market, closest competitors and other purely business
management factors. Most importantly, this literature has grown on
previous investigations of patterns of retail agglomeration that
looked at retail diversification and at spatial features
influencing rejection and attraction between categories of outlets
\cite{1987Brown,2007Coe,2006Jensen}. In this context, the
diversification of retail activities in an urban area is
considered as an important aspect of the social ``attractiveness''
of a city, and it is becoming nowadays subject of detailed
investigation.\cite{2014Carmona,2015Griffiths}.

Today, it is possible to study the location and diversity of
commercial activities in large real-world urban systems, by making
use of advanced GIS techniques and even mobile phone or
geo-located online social networks \cite{2011Noulas,
2012Porta,2009Porta,2014Ravulaparthy,2014Rui,2011Wang}.
Notwithstanding  the recent progress in studying cities as spatial
networks
\cite{BarthelemyPR2006,2006Boccaletti,2012SRstrano,2013SRstrano,2014JRSIlouf},
a quantitative analysis of the historical development of the
distribution of activities in a city is still missing, mainly
because of the difficulties in getting historical data in
electronic format. Building on a unique database, in this work we
contribute a first empirical study of the development of retail
activities in a city in time, by looking at the type and date of
birth of all the activities present at year 2004 along with their
geographical position. Specifically, the data set we have analysed
contains the exact position in space (location), the year of
registration and the market category (type) of each of the $N =
35,053$ commercial activities present in the city of Rome (Italy)
at the year 2004. Notice that, since the first activity was
registered on the $1^{st}$ of January 1900, while the last one on
the $1^{st}$ of July 2004, the entire data set covers a period of
more than a century. However, in our study we do not have complete
information on the whole set of activities present in Rome at a
certain time $t$, but only on those activities present at time $t$
which survived up to year 2004. In short, of all activities that
have populated the city of Rome in the analysis period, our data
set only refers to the ``survivors" at 2004, and we look backward
to their behaviors in time as associated to their type and
location. Each activity in fact belongs to one of eight commercial
categories (or types). In Table~\ref{t:categories} we report a
list of the categories together with the number of activities
$N^{\alpha}$ of each type $\alpha$, with $\alpha=1,2,\ldots,8$.

\begin{table}[ht]
\caption{Number $N^{\alpha}$ of commercial activities of type
$\alpha$ in Rome at year 2004} \centering
\begin{tabular}{c l r }
\hline\hline & Type $\alpha$ & Number $N^{\alpha}$
\\ [0.2ex] 
\hline 

 1  & Not Specialised (not alimentary) & 1946 \\
 2  & Food                &  3898 \\
 3  & Medicals            &  1640  \\
 4  & Goods               & 16225 \\
 5  & 2$^{nd}$ hand       &    130 \\
 6  & Unconventional (mail retail, street selling) & 4907 \\
 7  & Repair         & 1166 \\
 8  & Other         & 3025 \\

\end{tabular}
\label{t:categories}
\end{table}

\section*{Results}

\subsection*{Double trend of temporal growth}

We first looked at the temporal evolution of the number of
activities which survived up to 2004. We considered both the total
number of activities $N(t)$ {\em present} at time $t$ and still
active at time 2004, and the number of new activities $n(t)$ which
were {\em registered} at time $t$ and survived up to 2004. These
two quantities are related through the expression $N(t) =
\sum_{\tau=t_0}^t n(\tau)$ where $t_0$ is equal to year 1900. More
generally, in the following we will always use uppercase letters
for quantities related to the total number of activities $N(t)$
and lowercase letters for quantities related to the differential
number of activities $n(t)$.

\begin{figure}[t]
 \centering
\includegraphics[angle=-90,width=6.6cm]{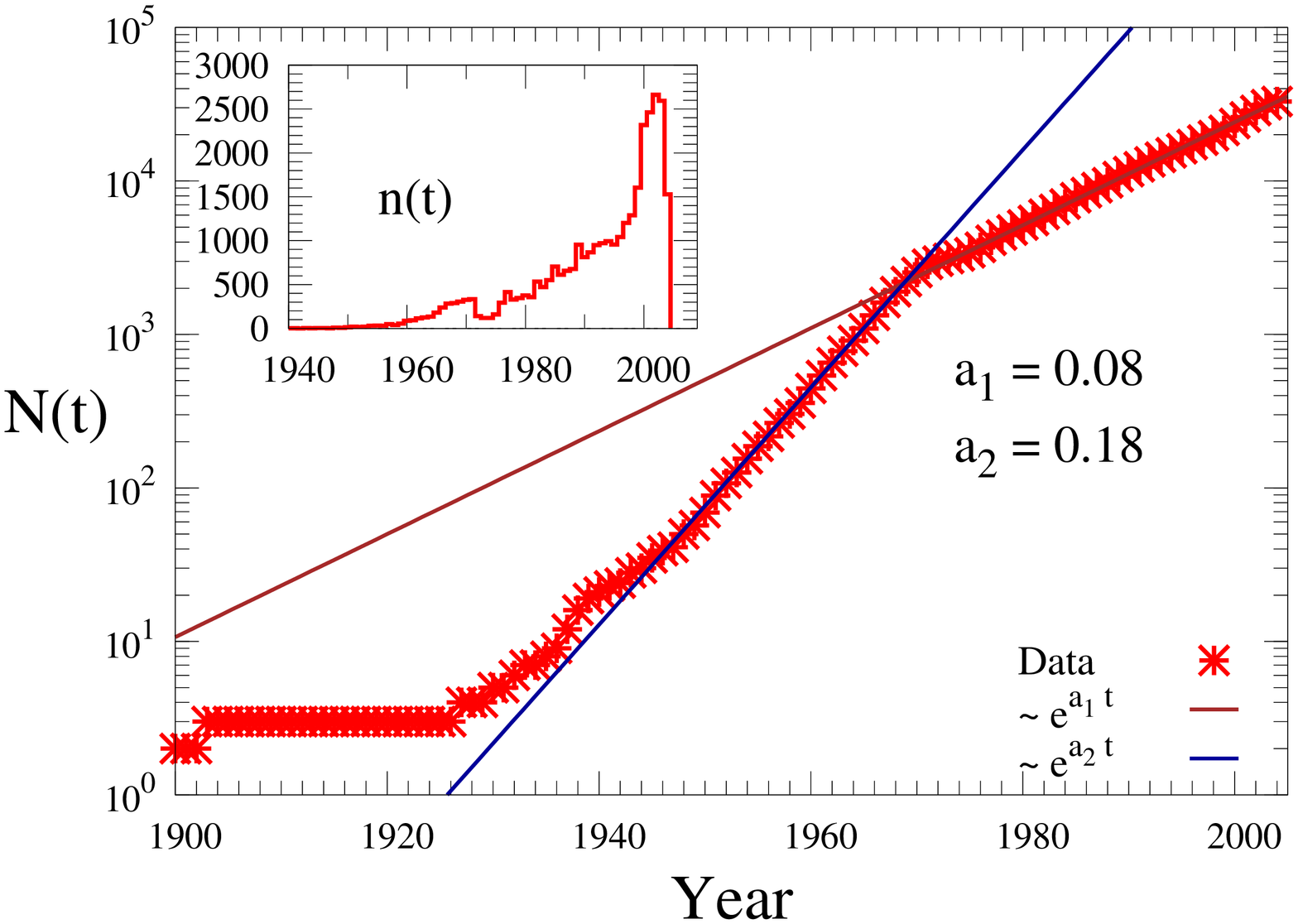}
\includegraphics[angle=-90,width=6.6cm]{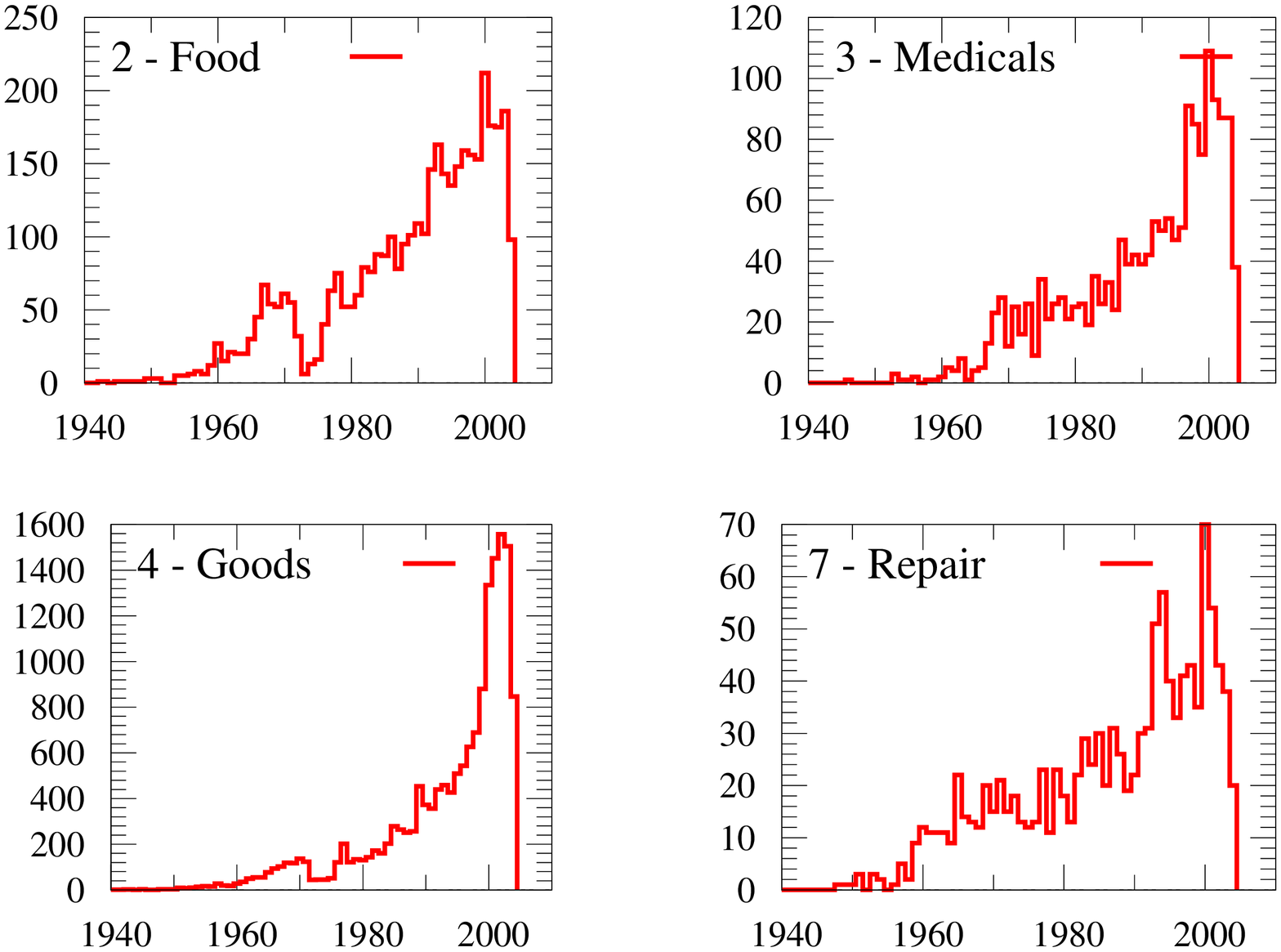}
\caption{{\bf Temporal evolution of the number of commercial
    activities.} Left: The cumulative number of survived activities, $N(t)$,
  i.e. the number of activities present at time $t$ in Rome, is
  plotted as a function of time $t$ (expressed in calendar years).
  The inset shows the number $n(t)$ of new registered activities at
  each year.  Right: The number $n^{\alpha}(t)$ of new activities is
  reported as a function of the year $t$, for a selection of the
  commercial categories reported in Table \ref{t:categories},
  i.e. Food, Medicals, Goods and Repair.}
 \label{Nk_C}
 \end{figure}

In the left panel of Fig.~\ref{Nk_C} we report the number $N(t)$
of activities present at time $t$ as a function of the year $t$ in
a semi-logarithmic plot. We notice the presence of two
well-defined exponential increases of the form $N(t) \sim e^{ a
t}$, characterised by different values of the parameter $a$,
respectively $a_1=0.08$ and $a_2=0.18$. The change of slope in
$N(t)$ occurs around the year 1975. In order to explain this
double trend, we report in the inset the number
$n(t)$ of activities registered at each year $t$ as a function of
$t$, in a double-linear scale. We observe that the number of
registered activities $n(t)$ exhibits a sudden drop in the period
1973-1975. That time interval coincides with the first oil crisis,
caused by the Yom Kippur Arab Israeli war and the consequent
embargo declared by the Organisation of Arab Petroleum Exporting
Countries (OAPEC) in October 1973, which had long-term effects not
just on the price of petroleum but also, and more importantly, on
all the major Western economies.
The right four panels of Fig.~\ref{Nk_C} report the number
$n^{\alpha}(t)$ of new activities registered each year for some
relevant commercial categories, i.e. ``Food", ``Medicals",
``Goods" and ``Repair" (the behaviour of $n(t)$ for the other
categories is qualitatively similar). Notice that a drop in the
number of new activities between 1973 and 1975 is visible, for
instance, for ``Food" and ``Goods", while other categories like
``Medicals" and ``Repair" do not show any sensible variation of
the overall increasing trend. This first result is a tangible
indication of how much the first oil crisis might have affected
the increase of commercial activities in Western Europe. It
appears that the crisis produced both a measurable decrease in the
number of new activities, and a long-term change in the increase
of the total number of activities after 1975, as testified by the
sharp change of slope of $N(t)$ in the left panel of
Fig.~\ref{Nk_C}.

\begin{figure}[t]
 \centering
\includegraphics[angle=-90,width=6.6cm]{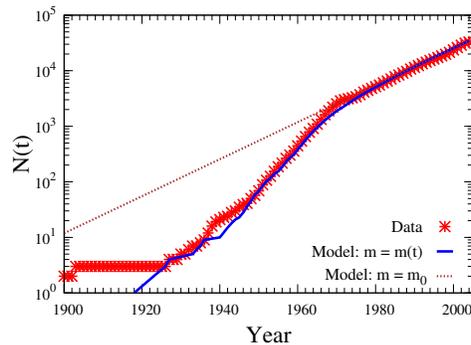}
\caption{{\bf Modeling the double trend.} The cumulative number
$N(t)$ of activities at time $t$ that survived at year 2004
obtained by the model of birth-death described in the text is
compared to the real data (symbols). The model produces an
exponential distribution (dashed-line) when the number $m$ of new
activities added every year is constant, while gives rise to the
empirically observed double trend when $m$ is exponentially
increasing in time until the year 1973 (full line).}
 \label{Nmodel}
 \end{figure}

In order to provide a quantitative explanation of the possible
origins of the double trend behavior reported in Fig.~\ref{Nk_C}
we have considered the following model for the temporal evolution
of the number of activities. The model is based on two mechanisms,
the death of some of the existing commercial activities and the
arrival of new ones. Namely, every year, starting from the year
1900, each of the existing activities dies with a probability $p$,
while $m$ new activities are added. This model can be solved
analytically and produces an exponential distribution for the
number of activities $N(t)$ present at time $t$. The distribution
obtained for $m=2500$ and $p=0.077$ is reported in
Fig.\ref{Nmodel} as dashed line and correctly reproduces only the
behavior observed in the city of Rome after the period 1973-1975.
If we want to capture the double trend found empirically, we need
to assume that the parameter $m$ of the model changes over time.
We have therefore assumed that $m$ increases exponentially with
the time, before the year 1973, while it stays constant in the
period after the crisis. This is justified by the increasing
prosperity of the city during the XIX century up to 1973. As for
the growing rate parameter $\alpha_m$ in the function $m(t)
\propto e^{\alpha_m t}$ we have used $\alpha_m = 0.11$. The
results of the numerical simulations of the model in this case are
reported in Figure \ref{Nmodel} as full lines. Notwithstanding its
simplicity, the model takes into account of the long lasting
impact generated by the crisis in the city of Rome, and results in
very good agreement with the data.

\subsection*{Diversification of activity types}

The attractiveness of an urban area is quite often related to the
diversity of resources made available to its inhabitants. When it
comes to commercial activities, such variety implies the presence
of retail shops belonging to several different categories. It is
therefore interesting to explore how the relative number of
activities of each category in our data set has evolved over time.
We start by defining the probability $p^{\alpha}(t) =
\frac{n^{\alpha}(t)}{n(t)}$ that a new activity of type $\alpha,
\alpha=1,\ldots,8$ is registered at time $t$, that is the ratio
between the number of new activities of type $\alpha$ at $t$ and
the total number of activities registered at time $t$.
Analogously, we define the probabilities $P^{\alpha}(t) =
\frac{N^{\alpha}(t)}{N(t)}$, which refer to the total number of
activities at time $t$.

\begin{figure}[t]
 \centering
\includegraphics[angle=-90,width=8.5cm]{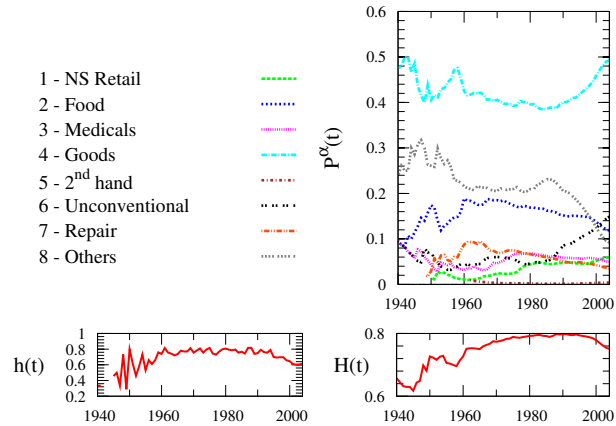}
\caption{{\bf  Time evolution of category diversification.}
Relative number of activities of type $\alpha$ present at time $t$
(upper panel). Differential (left bottom panel) and cumulative
(right bottom panel) entropies of category distributions are
reported as a function of time $t$.}
 \label{f:H}
 \end{figure}

In the upper panel of Fig.~\ref{f:H} we report the time evolution
of $p^{\alpha}(t)$ for each category $\alpha$, starting at year
1940 (the points before $t=1940$ were omitted from the plot due to
the lack of sufficient statistics).  Interestingly, even if the
total number of activities keeps growing, as illustrated above,
the relative abundance of categories remains approximatively
constant during a long period of about 30 years, between 1960 and
1990. A reorganisation of the relative abundance of activities is
instead observed in the period 1990-2004, and is mainly due to the
drastic increase in the number of ``Unconventional" activities and
the concurrent decrease in the total number of activities in
``Other".

We used two measures to concisely quantify the heterogeneity of
the activity distribution, namely the {\em category entropy}
$h(t)$, based on the registered activities at each time $t$, and
the {\em cumulative category entropy} $H(t)$. The first quantity
is defined as:
 \begin{equation}
  h(t) = -C_h \sum_{\alpha=1}^8 p^{\alpha}(t) \ln p^{\alpha}(t)
 \label{Hrt}
 \end{equation}
where $C_h=1/\ln{8}$ is a normalisation factor which guarantees
that $h(t)$ takes values in $[0,1]$.  The definition of $H(t)$ is
analogous, but is based on the cumulative distributions
$P^{\alpha}(t)$. The two entropies defined above can be used to
characterise the overall variety of activities present in the
city, ignoring the actual locations of retail shops (we will focus
on spatial distributions of activities in the next Section). In
particular, $h(t)=0$ only if we have exactly one type of
commercial activity, while we obtained $h(t)=1$ when
$p^{\alpha}(t)$ is a uniform distribution in the space of
categories, i.e. when there are exactly the same number of
activities for each category.

The values of $h(t)$ and $H(t)$ are shown in the two bottom panels
of Fig.~\ref{f:H}. After an initial phase characterised by a noisy
increase of their values, both entropies reach a stable plateau
which lasts for about three decades (roughly, between 1960 and
1990) and includes the overall maximum of each quantity over the
whole data set. The presence of a plateau around a maximum value
of entropy suggests that the attainment of a relatively
well-balanced variety of activities after the WWII, corresponding
with the economic growth in the 1950's and 1960's, has indeed
evolved into a stable pattern that has lasted for a considerably
long temporal interval, at least until the early 1990's, in
agreement with the stability of the relative abundance of
activities observed in the top panel of the same Figure.

It is interesting to notice that both $h(t)$ and $H(t)$ have
started decreasing in the mid-1990's, revealing that the
concentration of more recent activities is substantially different
from that that observed between 1960 and 1990. Specifically, we
can see that in the upper panel of Fig. 2 both ``Goods'' and
``Unconventional'' register a big increase, while ``Other'' and
``Food'' register an evident decrease. More generally, we observe
a first growth phase of the survived retails and their
diversification just after the war (urban expansion), followed by
a period of retail growth and diversity stability (urban
maturity), and then finally by a last decade of much slower retail
growth characterized by loss of diversity (urban stagnation) up to
the year 2004 where the entropy values have not yet stabilised to
a new equilibrium.

This decrease is possibly due to market distribution innovations
such as the spread of large scale retail followed by emerging
online styles of purchase. In addition to this, another possible
explanation is related to the occurrence of another factor, namely
the economical crisis of the Eighties in Italy, that reduced the
purchasing power of the Italian working class, and, possibly, the
variety of retail requests.

\subsection*{The spatial organization of commercial activities}

\begin{figure}[tbp]
 \centering \iftrue
 \includegraphics[angle=-90,width=6.6cm]{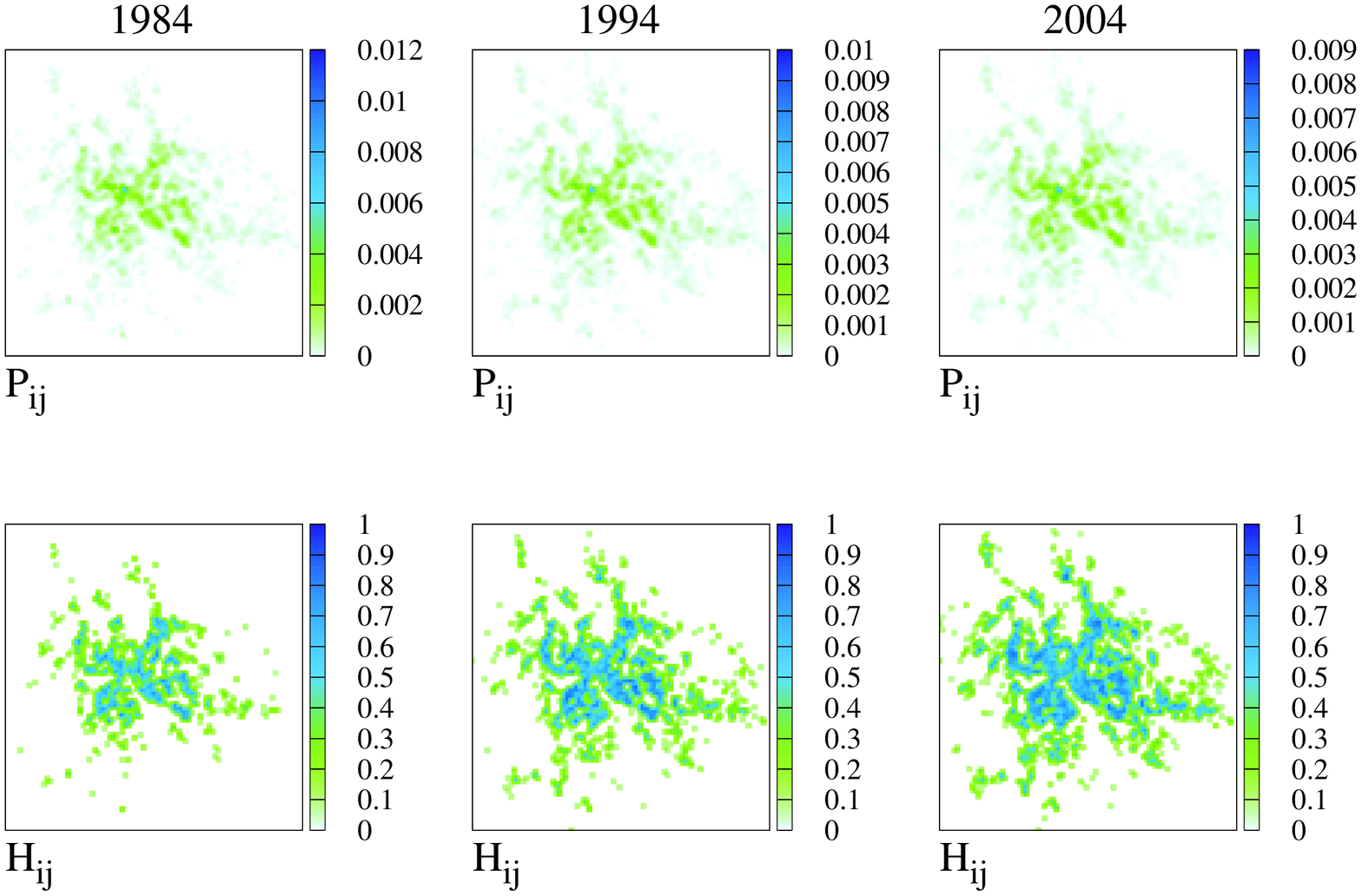}
 \includegraphics[angle=-90,width=6.6cm]{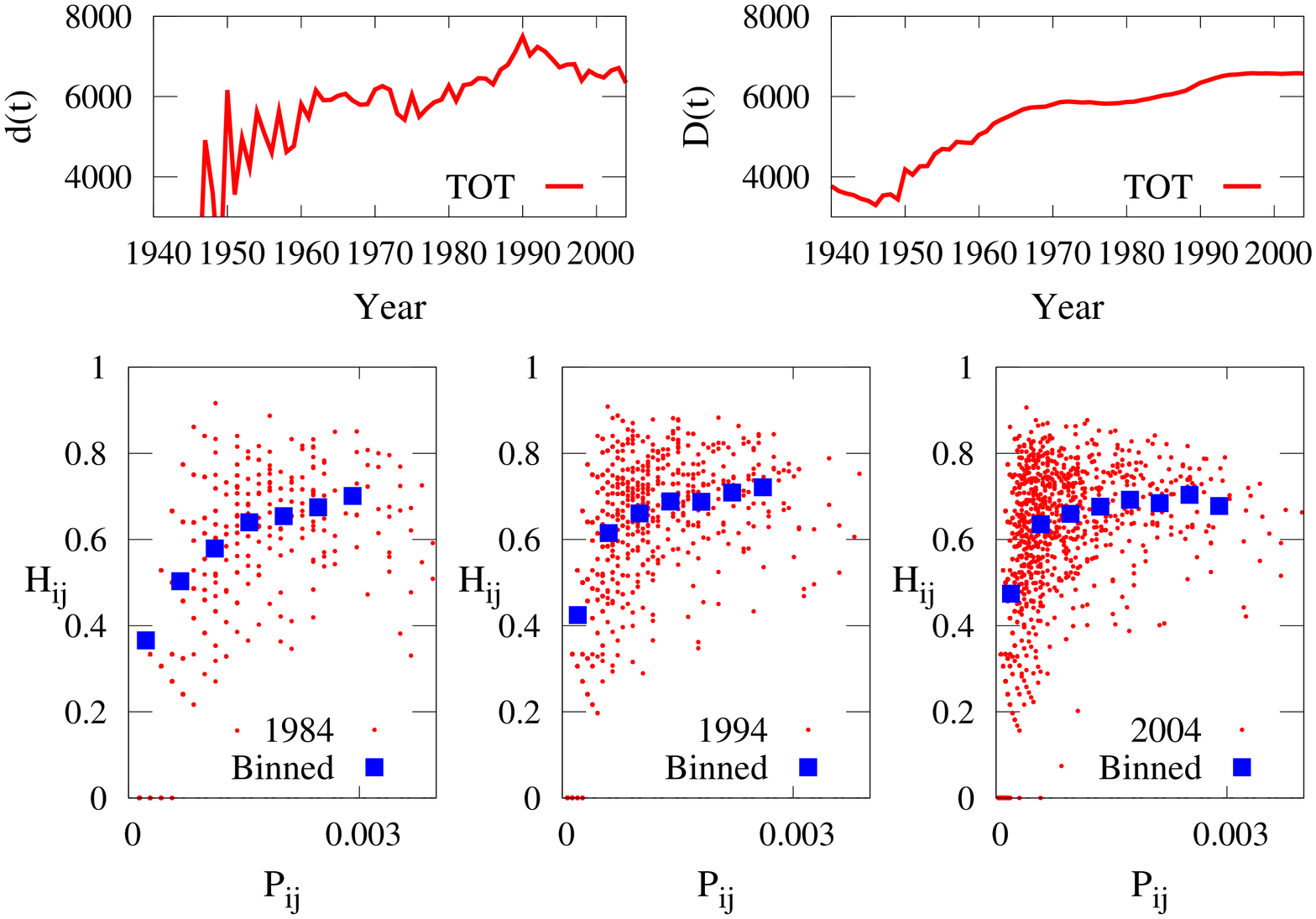} \else
  \vskip 6. truecm
 \fi

  \caption{{\bf Spatial distribution of activities and category
      diversification.} Spatial density of commercial activities
    (left/top panels), and distribution of the local category entropy
    $H_{ij}(t)$ (left/bottom panels) evaluated for the years 1984,
    1994, and 2004 ($1^{\rm st}$, $2^{\rm nd}$, and $3^{\rm rd}$
    column, respectively. We have considered a grid of $m \times m$
    cells, with $m=100$, corresponding to cells of linear size of
    $350\,{\rm m}$. The two upper/right panels show the square root of
    the mean square displacement from the city centre of all the
    activities, respectively in the registered ($d(t)$) and cumulative
    ($D(t)$ cases. Panels on the right/bottom show the scatter plots
    of density, $P_{ij}(t)$, {\it vs} entropy, $H_{ij}(t)$ for the
    same three years. }
 \label{f:Space_hc}
 \end{figure}

An important element that characterises urbanization is how
commercial activities and services are distributed in space. We
considered a uniform grid of $100\times 100$ square cells, each of
length $350$m, which covers all the activities present in the city
council of Rome in 2004, and registered over the previous century.
Each cell is identified by a pair of integer indices $(i,j)$, with
$i, j = 1 \dots 100$. We denote by $P_{ij}(t)$ the fraction of
activities present at time 2004 which where present at time $t$
inside the cell $(i,j)$, and by $p_{ij}(t)$ the fraction of new
activities registered at time $t$ which fall within cell $(i,j)$.
In the upper-left panels of Fig.~\ref{f:Space_hc} we show in a
density plot the distribution $P_{ij}(t)$ at three points in time,
namely in 1984, 1994, and 2004. It is clear that the spatial
distribution of activities in the urban area has considerably
evolved, with several zones becoming denser over time. Remarkably,
the total area of the city in which we find retail activities has
increased over time, as made evident by the plot of:
\begin{equation*}
  d(t) = \sqrt { \sum_i [{\bf r}_i(t) - {\bf r}_{CM}(t)]^2 }
\end{equation*}
reported in the top-right panels of Fig.~\ref{f:Space_hc}. Notice
that $d(t)$ is the square root of the mean square displacement of
the activity distributions at year $t$ with respect to their
centre of mass. The summation is extended over all the activities
registered at time $t$, ${\bf r}_i \equiv (x_i, y_i)$ is the
position of activity $i$, and ${\bf r}_{CM}(t) = \sum_i {\bf
r}_i(t) $. We report also the temporal evolution of $D(t)$, the
analogous quantity computed considering the total number of
activities present at time $t$.

We observe that, at the beginning of the past century, the activities
present in Rome were confined within a circle whose radius was about 3
km. The behavior of $D(t)$ is less affected by fluctuations and
clearly shows the important geographical expansion of the city during
the reconstruction period that followed the Second World War.  Such an
expansion lasted until the beginning of the 1970's, and then gave rise
to a period of 10-15 years when both $d(t)$ and $D(t)$ remained
constant. A second relevant expansion of the city is visible in the
decade 1985-1995, followed by a final stabilization lasting until
2004. The same trend is also observed when the quantity $D(t$) is
computed for each of the different commercial categories (results not
shown).

To better investigate the spatial distributions of the
diversification of activities in Rome we have also considered the
so-called \emph{local category entropy}. If we denote by
$p_{ij}^{\alpha}(t)$ the fraction of new activities within the
cell $(i,j)$ at time $t$ which are of type $\alpha$, the local
category entropy of the cell $(i,j)$ is defined as:
\begin{equation}
  h_{ij}(t) = -C_h \sum_{\alpha=1}^8 p_{ij}^{\alpha}(t) \ln
  p_{ij}^{\alpha}(t)
 \label{Hrc}
\end{equation}
and quantifies how balanced is the distribution of categories in
the cell. An analogous definition can be obtained from the
cumulative distributions $P_{ij}^{\alpha}(t)$, and is denoted in
the following as $H_{ij}(t)$.

The three panels in the bottom-left side of Fig.~\ref{f:Space_hc}
report as a colour-plot the values of $H_{ij}(t)$ for each cell,
respectively in 1984 (left), 1994 (middle), and 2004 (right). In
general, we observe a form of urbanization that proceeds
establishing local centres with higher diversity of retail around
which places with lower diversity contextually emerge as a ``grey"
background area. That seems to happen across scales, in a modular
fashion which is reminiscent of long established theories of urban
morphology~\cite{Caniggia2001}.

The three panels on the bottom/right of Fig.~\ref{f:Space_hc}
report the scatter plots of $P_{ij}(t)$ {\it vs} $H_{ij}(t)$,
again for the years 1984, 1994, and 2004. These plots exemplify
the relationship between accumulation and diversification in the
city, since $P_{ij}(t)$ and $H_{ij}(t)$ are measuring,
respectively, the concentration of commercial activities in a cell
and the heterogeneity of the categories in the same cell. Albeit
the two measures are somehow correlated, it is interesting to
notice that, for low values of $P_{ij}$, the corresponding values
of $H_{ij}$ increase almost linearly while, for higher densities,
the local category entropy saturates, supporting the idea that a
high diversification of categories requires a critical density of
activities to be achieved. The higher is the number of cells with
large values of entropy $H_{ij}$, the wider are the opportunities
locally available to city users.

 \begin{figure}
 \centering
\includegraphics[angle=-90,width=8.5cm]{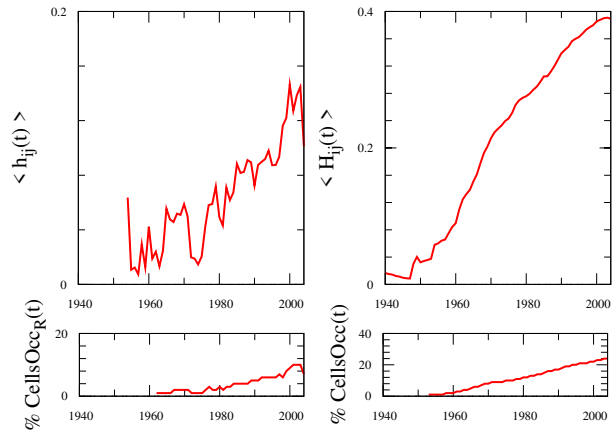}
\caption{{\bf Local properties.} The local activity entropy averaged
  over all the cells of the spatial grid (upper panels) in its
  differential (left) and cumulative version (right). Lower panels:
  Percentage of number of occupied cells of the grid.}
 \label{f:Averageh}
 \end{figure}

A more compact information can be extracted from
Fig.~\ref{f:Averageh}, whose upper panels show the average values
$\langle h_{ij}(t) \rangle$ and $\langle H_{ij}(t) \rangle$ of
local category entropies, where the averages are computed over all
the cells of the grid.  Interestingly, both these entropies do not
show the saturation observed in Fig.~\ref{f:H} for $h(t)$ and
$H(t)$, but are instead characterised by an overall increasing
trend over the whole period of interest, revealing that different
levels of diversification took place at both global and local
scales. Notice that the effects of the first oil crisis are again
visible in the plot of $h_{ij}(t)$ as a pronounced drop between
1973 and 1975, suggesting that the crisis affected the
distribution of retail activities at different spatial scales.
Finally, the lower panels of Fig.~\ref{f:Averageh} show the
percentage of the cells occupied by, respectively, at least one
new registered activity (left) and at least one activity (right).
Even in this elementary measure, the Oil Crisis period results
evident as a slight gap on the left plot and a small plateau on
the right one.

\subsection*{The network of attraction-repulsion between categories}

\begin{figure}
 \centering
\includegraphics[angle=-90,width=12.75cm]{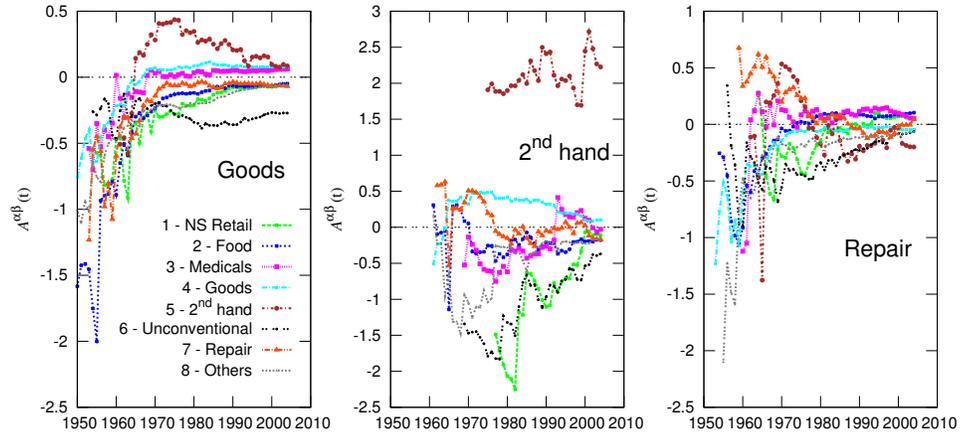}
\caption{{\bf Time evolution of attraction between categories.} The
  three panels show the values of the attraction coefficients ${\cal
    A}^{\alpha\beta}(t)$ respectively for $\alpha=4$ (``Goods''),
  $\alpha=5$ (``2$^{nd}$ hand''), and $\alpha=7$ (``Repair''), with
  respect to all the other categories, namely $\beta= 1,\ldots,8$. The
  radius $R$ was set equal to 200 meters.}
 \label{f:AR_JC}
 \end{figure}

\begin{figure} [tb]
 \centering
\includegraphics[angle=0,width=12.5cm]{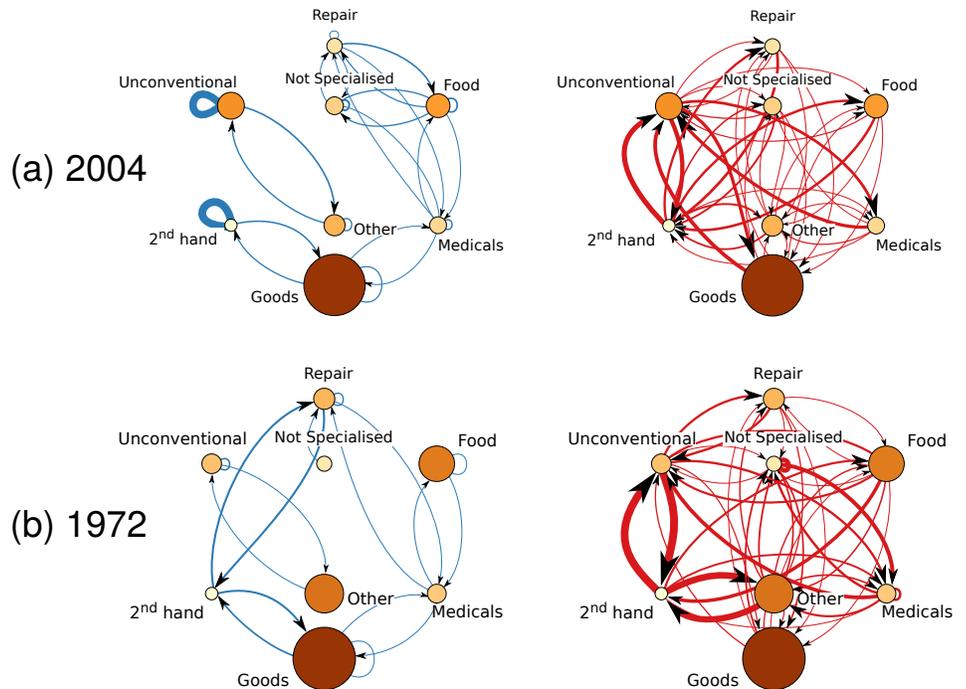}
\caption{ {\bf Networks of attraction and repulsion among activity
types.} The networks obtained from the spatial distribution of
activities at year 2004 (panel (a)), and at year 1972 (panel (b)).
The width of link  $\alpha \rightarrow \beta$  from category
$\alpha$ to category $\beta$ represents the value of $|\cal
A^{\alpha, \beta}|$.  Blue links indicate attractions (networks on
the left hand side), while red links indicate repulsions (networks
on the right hand side). The dimension of the nodes is
proportional to the fraction of commercial activities in the
corresponding category.
 }
 \label{f:N-AR-2004}
 \end{figure}

Important information on the spatial distribution of activities in
a city can be extracted by looking at how different commercial
categories attract or repel each others, and how such
attraction/repulsions patterns have evolved over the years.  As
proposed by Pablo Jensen in Ref.~\cite{2006Jensen}, a nice way to
capture the correlations between the spatial position of
commercial activities in a city is to construct a network where
each node represents a different commercial category, and where
the links stand for attraction/repulsion in space between couple
of categories. We have constructed such networks for each year
$t$, starting from the spatial distribution of activities, in the
following way. Given two activity types, namely $\alpha$ and
$\beta$, for each activity $i$ of type $\alpha$, we consider a
circle of radius $R$ centered at the position ${\bf r}_i=
(x_i,y_i)$ of $i$. Being $N_i^{\beta}(t)$ and $N_i(t)$
respectively the number of activities of type $\beta$ and the
total number of activities inside the circle centered in ${\bf
r}_i$ at time $t$, we consider the ratio $N_i^{\beta}(t) /
N_i(t)$.  We then compare $N_i^{\beta}(t) / N_i(t)$ to the
expected value of the ratio, $N^{\beta}(t) / N(t)$, that we would
obtain if the commercial activities of type $\beta$ were
distributed uniformly in space, independently from the positions
of the activities of type $\alpha$. If $N_i^{\beta}(t) / N_i(t)$
is different from $N^{\beta}(t) / N(t)$, this means that the
presence of the activity $i$ of type $\alpha$ at position ${\bf
r}_i$ affects the presence of activities of type $\beta$ in its
vicinity. We hence define the {\em attraction
  coefficient} ${\cal A}^{\alpha\beta}(t)$ between type $\alpha$ and
type $\beta$ at time $t$ as the logarithm of the ratio between
$N_i^{\beta}(t) / N_i(t)$ and $N^{\beta}(t) / N(t)$, averaged over all
circles centered around the $N^{\alpha}(t)$ activities of type
$\alpha$.  We finally get the following expression \cite{2006Jensen}:
 \begin{equation}
  {\cal A}^{\alpha\beta}(t) = \ln \left[ \frac{1}{N^{\alpha}(t)} \frac{N(t)}{N^{\beta}(t)} \sum_{i=1}^{N^{\alpha}(t)}
  \frac{N_i^{\beta}(t)}{N_i(t)} \right]
 \label{MabJc}
 \end{equation}
In this way, positive values of the attraction coefficient ${\cal
A}^{\alpha\beta}(t)$ indicate that the local density of activities
$\beta$ inside circles of radius $R$ centered at activities of
type $\alpha$ is higher than the average, and so category $\alpha$
attracts $\beta$. Conversely, negative values of ${\cal
A}^{\alpha\beta}(t)$ mean that $\alpha$ repels $\beta$, being the
density of $\beta$ inside circles centered at $\alpha$ lower than
the average. In the three panels of Fig.~\ref{f:AR_JC} we plot the
attraction coefficients of three categories, namely ``Goods'',
``2$^{nd}$ hand'', and ``Repair'', with respect to each of the
other eight categories as a function of time, where we considered
$R=200$ meters.
The global pattern of attraction/repulsion at a given year can be
represented and visualised in a graph. In
Fig.~\ref{f:N-AR-2004}(a) we report the attraction (left) and
repulsion (right) graph corresponding to year 2004. Each node
represents a commercial category, and its size is proportional to
the fraction of activities belonging to the corresponding
category. The directed link from node $\alpha$ to node $\beta$
indicates attraction (solid blue lines) or repulsion (dashed red
lines) between the corresponding categories, and the width of an
edge is proportional to the absolute value of the attraction
coefficient $\mathcal{A}^{\alpha \beta}$.

We found that in most of the case, except for the few self-loops,
attraction edges are relatively weak, and in small number, while
repulsion is usually much stronger, for instance for the pairs
``2$^{nd}$ hand-Unconventional'', ``Goods-Unconventional'', and
``Medicals-Unconventional''. Interestingly, several categories are
characterised by self-attraction, as demonstrated by the presence
of many blue loops, and can be extremely strong, such as in the
case of ``Unconventional'' and ``2$^{nd}$ hand''.  Conversely,
there is no big repulsion loops. As a consequence, it seems that
in general commercial activities of a given type tend to attract
other activities of the same type. Although the attraction and
repulsion graphs are directed, we remark that in most of the cases
edges are reciprocated, meaning that if a link exists from a node
$\alpha$ to a node $\beta$, then there is also a link from $\beta$
to $\alpha$, and the two links have a comparable weight. It is
interesting to notice that generally we do not observe the
concurrent presence of an attraction edge from $\alpha$ to $\beta$
and a repulsion edge from $\beta$ to $\alpha$, although such
configurations are possible in theory.

As a comparison, we report in Fig.~\ref{f:N-AR-2004}(b) the
network of attraction/repulsion at year 1972, namely before the
oil crisis. We notice that some of the relationships have been
maintained over the years, possibly with small changes in their
weight, while some are very different from the 2004 network. One
example of relevant persisting relationship is that of the pair
``Unconventional-2$^{nd}$ hand'', which was repulsive in 1972 and
remained repulsive in 2004, though with slightly different
intensities.  In other cases some links are, instead, not present
at one year, but appear in the other; examples are the couples
``Not Specialised-Repair'', or ``Food-Medicals'' which attract at
the year 2004, but have a slight asymmetry in the behavior in
1972, where ``Not Specialised'' attracts ``Repair'', but,
curiously, ``Repair'' repulse ``Not Specialised''.

It is interesting to notice that some attractive relationships
might turn into repulsive ones over time and vice-versa.  A
typical example is the relation between ``Repair-2$^{nd}$ hand''.
In fact, we can see from the third panel of Fig.~\ref{f:AR_JC}
that ``Repair'' attracts ``2$^{nd}$ hand'' before the year 1970,
while the opposite occurred after 1970. This behavior is also well
visible in the network representation of Fig.~\ref{f:N-AR-2004},
where we can see in panel (b) left of year 1972 a very clear blue
line between the two categories, while in the year 2004 the links
between the couple are red (panel (a) right). A similar situation
occurs for the couple ``Not Specialised-Food'' with a change from
weak repulsive to attractive around the same year.

The same is also valid for some self-attractions. An example is
given by the category ``Goods'', which was self-repulsive before
the 1970s, becomes self-attractive after 1970 (see the first panel
of Fig.~\ref{f:AR_JC}), and the category ``Not Specialised'',
which flipped from a clear self-repulsion into a weak
self-attraction around 1980. Table~\ref{t:changes} reports the
most important changes in the sign of pairwise relationships
observed in the data set. It is there evident that most of the
changes take place in one of three main periods, namely between
the end of the 1960s and the early 1970s, around the middle of the
1980s, and around year 1993, when the changes have been weak but
quite persistent in subsequent times. Interestingly, most changes
are from repulsion to attraction. The opposite direction occurs
only for the commercial type ``Repair'', which was self-attractive
before the year 1988, and self-repulsive afterwards for a couple
of years, and very slightly self-attractive again in the last
years before 2004.

It is important to notice that the attraction coefficient we have
considered compares the local concentration of activities with
respect to the concentration in the whole city. Due to this
normalisation, the evaluation of the attractiveness at year $t$,
based on the spatial distribution at $t$ of only those activities
which survived up to year 2004, can give information of the
properties of all the activities at year $t$ (under the assumption
that non-surviving activities have spatial distributions similar
to those of the surviving activities).
%
\begin{table}[ht]
\caption{Summary of the most relevant changes in the sign of
attraction between categories. The changes are generally
symmetrical with respect to the two categories.} \centering
\begin{tabular}{c l c c }
\hline\hline $\alpha-\beta$ & & $\approx$ year & Change \\
[0.2ex] \hline
 1 - 2  & Not Specialised  - Food & 1970 & $- \rightarrow + $ \\
 2 - 1  & Food - Not Specialised & 1970 & $- \rightarrow + $ \\
 3 - 4  & Medicals - Goods & 1967 & $- \rightarrow + $ \\
 4 - 3  & Goods - Medicals & 1967 & $- \rightarrow + $ \\
 4 - 4  & Goods - Goods & 1966 & $- \rightarrow + $ \\
\hline
 8 - 8  & Other - Other & 1986 & $- \rightarrow + $ \\
 1 - 1  & Not Specialised - Not Specialised & 1983 & $- \rightarrow + $ \\
 7 - 7  & Repair - Repair & 1988 & $+ \rightarrow - $ \\
\hline
 1 - 7  & Not Specialised - Repair & 1993 & $- \rightarrow + $ \\
 3 - 3  & Medicals - Medicals & 1993 & $- \rightarrow + $ \\
 3 - 5  & Medicals - 2$^{nd}$ hand & 1993 & $- \rightarrow + $ \\
 5 - 3  & 2$^{nd}$ hand - Medicals & 1993 & $- \rightarrow + $ \\
 7 - 1  & Repair - Not Specialised & 1993 & $- \rightarrow + $ \\
\end{tabular}
\label{t:changes}
\end{table}

Fig.~\ref{f:AR_JC} also shows a slightly tendency towards the
reduction of the value of $| { \cal A} |$ in time, and its
stabilization to a relatively low value. In fact, the commercial
activities start to become more evenly distributed in the
territory, and only a weak coexistence or repulsion of different
categories can be registered. This behaviour cannot be attributed
to the increase of the number of activities, because, as visible
in Fig.~\ref{f:H}, the density of activities is approximately
constant in time and the small variations observed in the density
do not affect the values of attractiveness.

\section*{Discussion}

In this paper we investigate location and type of all the 35,000
retail activities present in the city of Rome in 2004, along with
their development in time over approximately one century, from
1900 to 2004. The aim of the study is to shed some light on the
temporal dynamics of retail as one of the most fundamental drivers
of urbanization.

In order to do so, we first look at the number of activities both
in terms of registrations and cumulative presence year-by-year. We
notice that while activities restlessly grow over the whole
period, they do so at two very different speeds, with a marked
change in speed emerging in the middle of the Seventies.
This change immediately follows the first oil crisis, with
``Repair'' and ``Medicals" being the only two types of activities
that do not show any drop in registrations during the crisis. This
tells the story of an abrupt shift in the conditions of the local
market from a state of expansion to one of stagnation subsequent
to global political moves.

We then look at the diversity of retail types first globally, i.e. the overall contribution of each type to the whole retail stock present in 2004 year by year over the period, and locally, i.e. their
contribution year by year to the stock located in each square cell
of 350 meters of edge of a virtual grid covering the whole city of
Rome. We find that while the number of activities grows with time,
their diversity manifests a threefold global behaviour: it grows in
the years of the post-war reconstruction and expansion, remains quite
stable after the Sixties for about three decades, and decreases in the last decade. Locally, however, we see a different scenario, where on
one hand the higher diversity of retail activities occurs where their
spatial concentration is higher, while on the other there does not
appear to be any visible plafond to the local growth of retail
diversity. The observed difference in the global and local
organisation of activities, with the diversity of activity types
continuing steadily to grow locally in a condition of global
stability, can be of great importance in understanding the capacity of cities to find smaller scale forms of organisation that would normally go unnoticed.

\section*{Acknowledgments}
We acknowledge support from the EPSRC project GALE, EP/K020633/1
and EP/K020552/1. V.L. and V.N. acknowledge support from the
Project LASAGNE, Contract No.  318132 (STREP), funded by the
European Commission.


%
%
%

\end{document}